\documentclass[manuscript]{emulateapj}

 \usepackage{apjfonts}

\slugcomment{to appear in the Astrophysical Journal}

\shorttitle{Shortest recurrence periods of novae}
\shortauthors{Kato et al.}

\begin{document}


\title{SHORTEST RECURRENCE PERIODS OF NOVAE}


\author{Mariko Kato} 
\affil{Department of Astronomy, Keio University, Hiyoshi, Yokohama
  223-8521, Japan;}
\email{mariko@educ.cc.keio.ac.jp}

\author{Hideyuki Saio}
\affil{Astronomical Institute, Graduate School of Science,
    Tohoku University, Sendai, 980-8578, Japan}

\author{Izumi Hachisu}
\affil{Department of Earth Science and Astronomy, College of Arts and
Sciences, The University of Tokyo, 3-8-1 Komaba, Meguro-ku, 
Tokyo 153-8902, Japan}

\and

\author{Ken'ichi Nomoto\footnote{Hamamatsu Professor}}
\affil{Kavli Institute for the Physics and Mathematics of the Universe 
(WPI), The University of Tokyo, 5-1-5 Kashiwanoha, Kashiwa,
Chiba 277-8583, Japan}




\begin{abstract}  
Stimulated by the recent discovery of the 1 yr recurrence period nova 
M31N 2008-12a, we examined the shortest recurrence periods of hydrogen
shell flashes on mass-accreting white dwarfs (WDs).  We discuss 
the mechanism that yields a finite minimum recurrence period 
for a given WD mass.  Calculating the unstable flashes 
for various WD masses and mass accretion rates, 
we identified a shortest recurrence period of about two months
for a non-rotating 1.38 $M_\sun$ WD with a mass accretion rate of 
$3.6 \times 10^{-7}~M_\odot$~yr$^{-1}$.  A 1 yr recurrence period
is realized for very massive ($\gtrsim 1.3~ M_\odot$) WDs with very
high accretion rates ($\gtrsim1.5 \times 10^{-7}M_\odot$~yr$^{-1}$). 
We revised our stability limit of hydrogen shell burning,
which will be useful for binary evolution calculations toward 
Type Ia supernovae. 
\end{abstract}




\section{INTRODUCTION}
\label{sec_introduction}

The recent discovery of the recurrent nova M31N 2008-12a
has attracted attention 
to novae with short recurrence periods \citep{dar14, hen14, tan14}. 
M31N 2008-12a showed the shortest recorded recurrence period of 1 yr, 
a very rapid turn-on of the stable supersoft X-ray source (SSS) phase, 
and a high effective temperature ($\sim 100$ eV) in the SSS phase, 
all of which indicate a very massive white dwarf (WD).  Such massive 
WDs in recurrent novae are considered to be one of the candidates for 
Type Ia supernova (SN~Ia) progenitors 
\citep{hku99, hkn99, hac01kb, hkn10, han04, li97, kat12review, pag14}. 
SNe Ia play very important roles in astrophysics as standard candles 
for measuring cosmological distances and as the main producers of iron 
group elements in the chemical evolution of galaxies.  However, their 
immediate progenitors just before SN Ia explosions are still unclear.
Thus, studies of novae with very short recurrence periods 
are essential to identify immediate progenitors of SNe~Ia.

Many theoretical works on hydrogen shell flashes have been published.  
In general, short recurrence periods are obtained for very massive WDs
with high mass accretion rates.  When the mass accretion rate 
exceeds a certain value, nuclear burning is stable,
and no shell flashes occur \citep{sie75, sie80, sio79, ibe82, nom07, wol13}.
The border between stable and 
unstable mass accretion rates is known as the stability line, 
i.e., $\dot M_{\rm stable}$, 
in the diagram of accretion rate vs. WD mass. 
For a given WD mass, the shortest recurrence period is obtained 
near the stability line. However, it is not well known whether this 
minimum recurrence period approaches a finite value or zero.  
\citet{wol13} recently obtained numerically the recurrence 
periods near the stability line for various WD masses and showed that 
the minimum recurrence period is finite. 
However, the theoretical reason for the finite value is still unclear. 
Moreover, the stability line obtained by \citet{wol13} using a 
time evolution calculation differs slightly from that obtained  
using a linear stability analysis \citep{sie75, sie80, nom07}. 
Because the stability line is important
in binary evolution calculations toward SNe~Ia, we examine the stability
line of shell flashes and clarify why there is a finite minimum value
of the recurrence period.

In the next section, we explain the reason for the finite minimum values of 
the nova recurrence period.  In Section \ref{sec_evolution}, we present
calculations of shell flashes on very massive WDs and numerically
obtain the minimum recurrence periods for various WD masses.   
We also present a recalculated stability line, which could be useful
for calculations of binary evolution.  In Section \ref{sec_discussion}, 
we discuss some numerical calculations that resulted in
shell flashes for mass accretion rates above the stability line. 
Finally, we summarize our results in Section \ref{conclusions}.


\begin{figure}
\epsscale{1.2}
\plotone{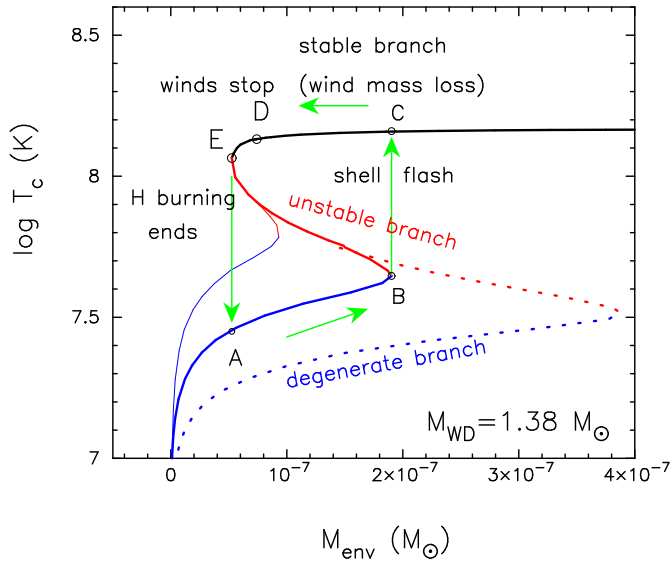}
\caption{
Schematic $M_{\rm env}-\log T_{\rm c}$ diagram for one cycle of nova
outbursts on a $1.38~M_\sun$ WD with a given mass accretion rate $\dot M$.
Here, $T_{\rm c}$ is the temperature at the bottom of the hydrogen-rich
envelope, and $M_{\rm env}$ is the envelope mass. 
We plot three mass accretion rates: above the stability
line, i.e., $\dot M > \dot M_{\rm stable}$ (thin solid line); somewhat
below the stability line, i.e., $\dot M < \dot M_{\rm stable}$
(thick solid line); and much below the stability line, i.e., 
$\dot M \ll \dot M_{\rm stable}$ (dotted line).
For the thick solid line, the WD starts accreting around point A
and the envelope mass increases.  When it reaches turning point B,
a shell flash begins. Thus, the ignition mass is defined as
$M_{\rm ig}= M_{\rm env}({\rm B})$.
Then the envelope expands and reaches point C. 
After an optical maximum at point C, the envelope mass decreases 
owing to wind mass loss and nuclear burning. The optically thick 
wind stops at point D. Hydrogen burning ends at point E, 
and the envelope quickly cools toward point A. In this S-shaped
sequence ABEDC,   
the lower branch (blue line from A to B) represents a degenerate envelope,
and the middle branch (red line from B to E) represents an unstable envelope
for nuclear burning.  The upper branch (black line from E to C) represents
an extended envelope after optical maximum, where nuclear burning
is stable.  For the lower mass accretion rate (dotted line),
the degenerate branch is cooler owing to the smaller gravitational 
energy release; hence, the ignition mass is larger.
For the higher mass accretion rate, 
the ignition mass is smaller.  We found that the width of 
this limit cycle, $\Delta = M_{\rm env}({\rm B}) - 
M_{\rm env}({\rm E})$, does not vanish even for very high 
mass accretion rates above the stability line (thin solid line).  
This is the reason for the minimum recurrence period of 
$t_{\rm rec}^{\rm min}= \Delta / \dot M_{\rm stable}$.
See text for details.
\label{flash}}
\end{figure}

\section{LIMIT CYCLE OF HYDROGEN SHELL FLASHES 
AND FINITE RECURRENCE PERIODS}
\label{section_limitcycle}

We first discuss the cycle of shell flashes using hydrostatic envelope
models.  We calculated the structures of the hydrogen-rich envelopes 
on a mass-accreting WD by solving the equations of hydrostatic equilibrium,
mass continuity, energy conservation, and energy transfer 
together with the equation of state for degenerate matter \citep{egg73}.
We used the OPAL radiative opacities \citep{igl96}.
The equation of energy conservation includes terms for nuclear burning, 
gravitational energy release, and radiative loss.  For the gravitational
energy release, we calculated the so-called homologous compressional heating, 
$\varepsilon^{\rm h}_g$, by accretion in a quasi-steady state 
on the $q$-coordinate ($q\equiv M_r/M$), which is proportional to 
the mass accretion rate $\dot M$ \citep[see, e.g.,][]{neo77, kat80, kat82}.

Figure \ref{flash} schematically depicts one cycle of shell flashes
in the $M_{\rm env}-\log T_{\rm c}$ diagram for a $1.38~M_\sun$ WD with
a given mass accretion rate of $\dot M$, where $M_{\rm env}$ is 
the envelope mass, and $T_{\rm c}$ is the temperature at the bottom 
of the hydrogen-rich envelope.  Here, we plot three mass accretion
rates: above the stability line, i.e., $\dot M > \dot M_{\rm stable}$
(thin solid line); somewhat below the stability line, i.e., 
$\dot M < \dot M_{\rm stable}$ (thick solid line);
and much below the stability
line, i.e., $\dot M \ll \dot M_{\rm stable}$ (dotted line).
Note that the exact stability line will be determined
by time-dependent calculation of the shell flashes 
in Section \ref{section_stability}. 
The hydrostatic solutions form an S-shaped sequence.
The lowest branch, dubbed ``the degenerate branch,'' 
represents a geometrically thin envelope in which 
most of the radiative
energy loss is balanced by gravitational energy release 
because very little energy is generated by nuclear burning. 
On the middle branch, dubbed ``the unstable branch,''
the energy loss is balanced by the energy generated by 
hydrogen burning and gravitational energy release 
($\varepsilon_g^{\rm h}$).  
The energy generation by nuclear burning is comparable to or even
dominates the gravitational energy release. 
Hydrogen shell burning on the middle branch is unstable.
The upper branch, dubbed ``the stable branch,'' represents an expanded
phase of the nova outburst. On the right side to point D, 
optically thick winds occur, and each envelope solution is constructed 
including wind mass loss \citep{kat94h}. 
Hydrogen shell burning is stable on this branch \citep{kat83}. 

For the thick solid line ($\dot M < \dot M_{\rm stable}$)
in Figure \ref{flash}, a mass-accreting WD begins its evolution
from somewhere on the degenerate branch around point A. 
The envelope mass increases with time because of accretion and
moves rightward along the degenerate branch. 
After it reaches the turning point of the degenerate branch (point B), 
the hydrogen shell burning becomes unstable, triggering a shell flash.  
The envelope structure changes from a geometrically thin configuration
to a very bloated one. The photosphere expands, and the photospheric
temperature rapidly decreases.  This is the onset of a nova outburst.
The envelope mass at point B corresponds to the ignition mass, i.e.,
$M_{\rm ig}=M_{\rm env}({\rm B})$. 

After the WD jumps to point C, it moves leftward as the envelope mass
decreases owing to wind mass loss and nuclear burning
\citep{kat94h, kov98, hac01kb}.
This ``stable branch'' corresponds to expanded envelopes in the 
decay phase of a nova outburst.  The photospheric radius gradually
shrinks while the photospheric temperature increases.  The optical 
brightness decays with time \citep{ibe82,kat94h, hac06kb}.
After the optically thick winds stop at point D, the envelope mass 
continues to decrease owing to nuclear burning.  As the envelope mass
decreases, the radius of the envelope continues to decrease, 
so that the geometry of the envelope gradually changes 
from a spherically, extended configuration to a thin one.
As a result, $T_{\rm c}$ decreases with the decreasing envelope mass
from point D to E.  When the WD reaches a turning point 
(point E), hydrogen burning becomes inactive because 
the envelope mass is too small to support high temperatures.
The unstable branch starts at point E. 
The nuclear burning rate at point E should be the same as 
that at the stability line, i.e.,
\begin{equation}
\dot M_{\rm stable} = \dot M_{\rm nuc}({\rm E}),
\label{stable_eq}
\end{equation}
where $\dot M_{\rm nuc}= L_{\rm n}/XQ$ is the nuclear burning rate, 
$L_{\rm n}$ is the luminosity of hydrogen burning, 
$X$ is the hydrogen mass fraction,
and $Q=6.4\times10^{18}$~erg~g$^{-1}$ is the energy
generation per unit mass.  The star quickly falls to point A 
on the degenerate branch. In this way, a nova follows a limit cycle of
points A, B, C, D, E, and then A.  

Figure \ref{flash} also shows two other tracks for lower
($\dot M \ll \dot M_{\rm stable}$, dotted line)
and higher ($\dot M > \dot M_{\rm stable}$, 
thin solid line) mass accretion rates.
For the lower mass accretion rate, the degenerate branch is cooler 
owing to the smaller gravitational energy release, but the position of
the unstable branch (from point E to B) hardly changes.
For this reason, point B is shifted rightward for the lower accretion rate; 
hence, the ignition mass is larger.  For the higher mass accretion rate, 
on the other hand, point B moves leftward.  As a result, the ignition mass
is smaller.  The higher the mass accretion rate is, the narrower 
the limit cycle of nova outbursts is.  

The width of this limit cycle $\Delta$ is defined for  
each S-shaped line as 
\begin{equation}
\Delta = M_{\rm env}({\rm B}) - M_{\rm env}({\rm E}),
\end{equation}
where $M_{\rm env}({\rm B})$ and $M_{\rm env}({\rm E})$ are
the envelope masses at turning points B and E, respectively. 
We found that the width 
does not vanish even for very high mass accretion rates above 
the stability line ($\dot M > \dot M_{\rm stable}$, 
thin solid line in Figure \ref{flash}). 
In other words, the finite $\Delta$ is the reason for 
the finite minimum recurrence period of
\begin{equation}
t_{\rm rec}^{\rm min}=  \Delta / \dot M_{\rm stable},
\end{equation}
when the mass accretion rate increases
toward the stability line, $\dot M \rightarrow 
\dot M_{\rm stable}$.  

\citet{ibe82} also presented the stability line  in the HR diagram  
as a locus of minimum envelope masses of accreting WDs.
The envelope mass has a minimum value at the knee near the highest
temperature on the locus of an accreting WD. This corresponds to 
point E in Figure \ref{flash}. \citet{ibe82} wrote that 
``the steady-burning solution is unstable along the low luminosity branch.''
\citet{fuj82} also discussed the stability change
at point E in relation to the change of envelope structure
from a spherical configuration (upper branch) to a flat configuration
(lower branch).
\citet{sal05} calculated the minimum envelope masses for various 
WD masses and chemical compositions (CO/ONe rich) using the OPAL opacities.
They wrote, ``Hydrogen burning will continue until the minimum 
envelope mass and maximum effective temperature are reached; evolution
cannot proceed further with stationary hydrogen-burning.''
Thus, hydrogen burning is extinguished, and the envelope enters
the cooling phase. 

The tracks in Figure \ref{flash} differ slightly from 
those of our time-dependent calculations, 
as will be described in detail in Section \ref{sec_evolution}.
We will identify the exact stability line using a 
time-dependent calculation in Section \ref{section_stability}. 

Finally, we discuss the case that the mass accretion rate is larger than
that of the stability line. For the thin solid line 
($\dot M > \dot M_{\rm stable}$) in Figure \ref{flash}, the WD accretes
hydrogen-rich matter and reaches turning point B. 
Hydrogen ignites to trigger a shell flash.  
After the WD jumps from turning point B to point C, 
it stays at a point on the stable branch to the right of 
point E because 
the mass accretion rate is larger than the nuclear burning rate
at point E; i.e., $\dot M > \dot M_{\rm nuc}({\rm E}) = \dot M_{\rm stable}$. 
We emphasize that, after the first shell flash occurs, the WD stays
at a point on the stable branch, and the shell flashes never repeat. 
We will discuss this in more detail in Section \ref{enforced_novae}.

\begin{figure}
\epsscale{1.2}
\plotone{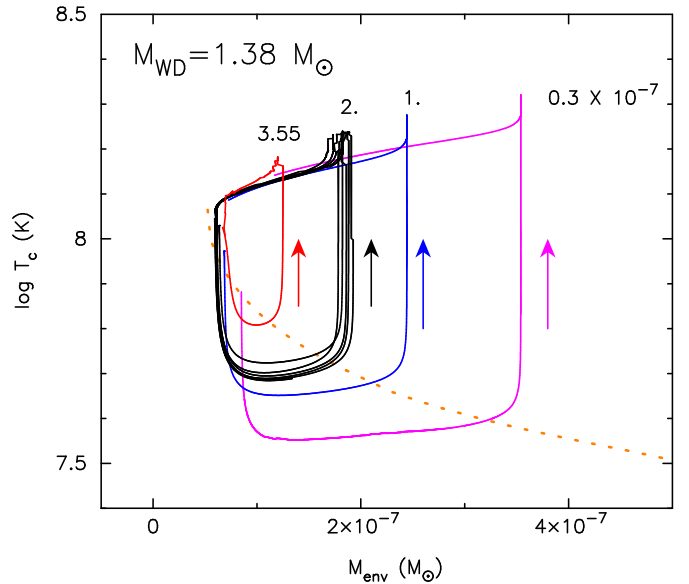}
\caption{
Same as Figure \ref{flash}, but for time-dependent calculation 
on a $1.38~M_\sun$ WD.  From left to right, the mass accretion rate is 
$\dot M=3.55 \times 10^{-7}$ (red solid line), 
$2 \times 10^{-7}$ (black solid line), $1 \times 10^{-7}$ (blue solid line), 
and $0.3 \times 10^{-7}~M_\sun$~yr$^{-1}$ (magenta solid line), 
as indicated next to each curve in units of $10^{-7}~M_\sun$~yr$^{-1}$.
For $2 \times 10^{-7}~M_\sun$~yr$^{-1}$,  
five cycles are shown, whereas for the other cases,
the first cycle is shown. 
We define the ignition mass $M_{\rm ig} = M_{\rm env}$ 
as the envelope mass at each vertical line of the first shell flash.
An orange dotted line shows the unstable branch in Figure \ref{flash}. 
\label{TcMenvM138X70.s}
}
\end{figure}

\section{SIMULATION OF SHELL FLASHES}
\label{sec_evolution}

In this section, we present and discuss the results of time evolution 
calculations for part of the nova cycle.  
We used a Henyey-type evolution code, 
which is the same as that in \citet{nom07} except that a term proportional to 
$(\partial s/\partial t)_q$, where $s$ is the specific entropy, 
is included to follow the evolution. 
The chemical composition of the accreting matter and initial
hydrogen-rich envelope 
is assumed to be $X=0.7, ~Y=0.28$, and $Z=0.02$, except for 
the cases in section \ref{sec_composition}. 
Neither convective overshooting nor diffusion processes are included; 
thus, no WD material is mixed into the hydrogen-rich envelope.

As the initial model, we adopted an equilibrium model 
\citep[i.e., ``the steady-state models'' of ][]{nom07}, 
in which an energy balance is already established
between heating by mass accretion
and nuclear energy generation and cooling by radiative transfer 
and neutrino energy loss.  This is a good approximation, as we will see
later, of the long time-averaged evolution of a mass accreting WD.
Thus, starting from such an initial equilibrium state, 
we expect that the ignition mass estimated on the basis of the 
first flash is close to the final one.

   Using this equilibrium model for the initial evolution model 
is advantageous because, after many flashes, the temperature distribution 
in the WD interior in a recurrent nova is expected to be similar 
to that of the equilibrium model accreting at the same rate.
This is a good approximation, in particular, for short-period recurrent
novae with very high accretion rates.  We assumed no helium layer between
the WD core and hydrogen-rich envelope. 
In some cases, we followed several shell flashes
to confirm that the equilibrium approximation is adequate.

Henyey-type evolution codes are well known to have difficulties
in calculating extended envelopes after a shell flash. 
To continue our calculation, we assumed large mass loss rates 
so that the radius does not exceed 0.01 -- 0.02~$R_\sun$. 
The mass loss rate thus determined is artificial and does not
represent any physical processes in nova outbursts. 
Although we adopted the smallest possible mass loss rates,
this artificial mass loss would shorten the duration of a flash
(i.e., the high state).  We suppose, however, that it has 
little effect on the resultant recurrence period because the durations 
of the flashes are much shorter than the recurrence period.

\begin{figure}
\epsscale{1.2}
\plotone{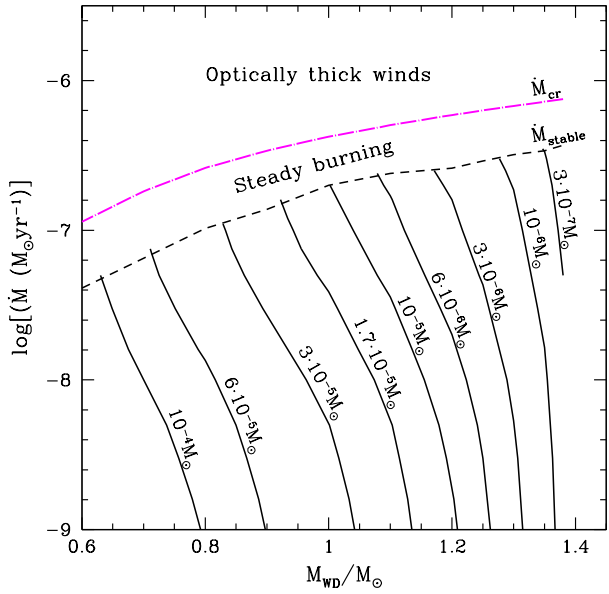}
\caption{
Ignition mass $M_{\rm ig}$
on the $M_{\rm WD}-\dot M$ plane.  Hydrogen burning is
stable in the region above the dashed line ($\dot M_{\rm stable}$). 
In the region below $\dot M_{\rm stable}$, 
hydrogen shell burning is thermally unstable,   
and the WD experiences shell flashes.  Black solid lines indicate 
equi-ignition masses, the values of which are shown beside each line. 
Optically thick winds are accelerated in the region 
above the dash-dotted line ($\dot M_{\rm cr}$). 
\label{revised_Nomoto_diagram}}
\end{figure}

\subsection{Ignition Mass} 
\label{sec_ignition_mass}
We obtain the ignition mass for a given WD mass and mass accretion rate 
as follows: We start our evolution calculation by reducing the envelope
mass to half that of the corresponding equilibrium model.
The nuclear luminosity $L_{\rm n}$ initially decreases   
because of the reduced envelope mass.   
This process corresponds to a falling path from a point on the unstable 
equilibrium branch toward the degenerate branch in Figure \ref{flash}.
As the envelope mass is increased by accretion, the WD moves rightward
and reaches point B, at which hydrogen burning becomes unstable, 
and a shell flash occurs.
We define the ignition mass $M_{\rm ig}=M_{\rm env}$
as the envelope mass at the onset of the flash. 
The ignition mass obtained in this way depends very little 
on the process or 
amount of initial reduction of the envelope mass; 
e.g., no reduction or a 90\% reduction yields 
a similar value of $M_{\rm ig}$.

Figure \ref{TcMenvM138X70.s} shows the evolutionary courses of 
these calculations
for a 1.38 $M_\sun$ WD in the $M_{\rm env} - \log T_{\rm c}$ diagram
for various  mass accretion rates.  
The orange dashed line represents the unstable branch in Figure \ref{flash}. 
After a steep decrease in $T_{\rm c}$ 
(and hence in the nuclear luminosity $L_{\rm n}$),
the envelope mass $M_{\rm env}$ increases to the ignition mass 
$M_{\rm ig}$, and a shell flash starts.
As the shell flash sets in, the temperature $T_{\rm c}$ and 
nuclear energy generation rate $L_{\rm n}$ 
increase rapidly, so the entire envelope expands. 
We define the ignition mass as the mass at which the temperature 
rises steeply in the first flash.  
Although the ignition mass changes slightly for every flash, as shown 
for the $2 \times 10^{-7}M_\sun$~yr$^{-1}$ case, 
the ignition mass at the first flash is sufficiently close 
to that of the limit cycle. This is one of the merits of starting
from the equilibrium model. 
The ignition mass thus determined will be examined in more detail
in Section \ref{multi_cycles}.

We obtained the ignition masses $M_{\rm ig}$ for various sets of mass 
accretion rates $\dot M$ and WD masses $M_{\rm WD}$.  
The results are presented as 
contours of $M_{\rm ig}$ on the $M_{\rm WD}-\dot M$ plane
in Figure \ref{revised_Nomoto_diagram}. 
For a given $\dot M$, $M_{\rm ig}$ is smaller for a larger $M_{\rm WD}$.
For a given  $M_{\rm WD}$, $M_{\rm ig}$ is smaller for a higher $\dot M$. 
All the contours stop at the dashed line because 
above this line, hydrogen shell burning is stable, and no flashes occur;
in other words, it is the stability line for hydrogen shell burning.
The stability line (dashed line in Figure \ref{revised_Nomoto_diagram}) is
approximately represented by 
\begin{equation}
 \dot M_{\rm stable}  = 4.17 \times 10^{-7} \left( {M_{\rm WD} 
\over M_\sun} -0.53 \right) ~M_\sun~{\rm yr}^{-1} .
\end{equation}
We will discuss how the stability line is obtained in the next subsection.

The dash-dotted line in Figure \ref{revised_Nomoto_diagram} represents
the locus of the critical accretion rate $\dot M_{\rm cr}$, above which
steady models have optically thick winds \citep{kat94h}.
This relation can be represented as
\begin{equation}
 \dot M_{\rm cr}  = 8.18 \times 10^{-7} \left( {M_{\rm WD} 
\over M_\sun} -0.48 \right) ~M_\sun~{\rm yr}^{-1} .
\end{equation}

\begin{figure}
\epsscale{1.2}
\plotone{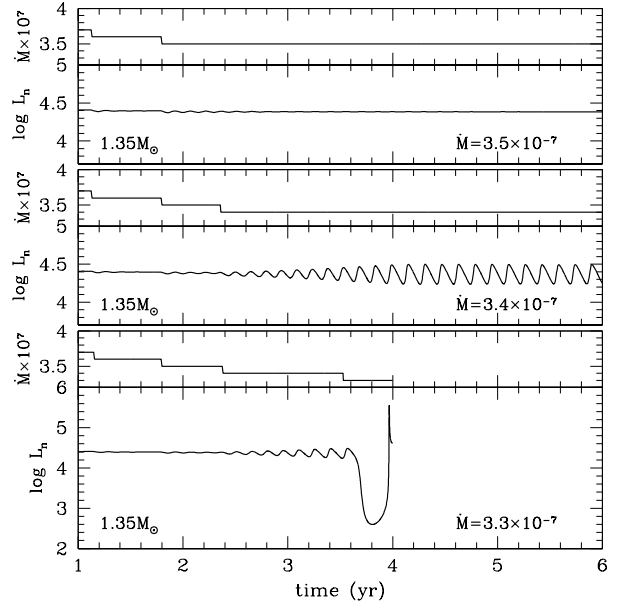}
\caption{ Evolutions of nuclear luminosity $L_{\rm n}$ 
(in  units of $L_\sun$) for selected accretion rates  
$\dot M$  (in units of $M_\sun$~yr$^{-1}$) around
the boundary between stable and unstable shell hydrogen burning
for a $1.35~M_\sun$ WD.
Each calculation started with the equilibrium model for an accretion rate
of $4.5\times10^{-7}M_\sun$~yr$^{-1}$, which is high enough for 
stable hydrogen shell burning. 
The accretion rate is gradually decreased (upper sub-panel of each panel) 
to $\dot M=3.5\times10^{-7}~M_\sun$~yr$^{-1}$ in the top panel,
$\dot M=3.4\times10^{-7}~M_\sun$~yr$^{-1}$ in the middle panel, 
and $\dot M=3.3\times10^{-7}~M_\sun$~yr$^{-1}$ in the bottom panel.
The stability of the shell burning changes at 
$\dot M=3.4\times10^{-7}~M_\sun$~yr$^{-1}$ (middle panel), 
at which the nuclear luminosity oscillates without leading to a flash. 
For an accretion rate slightly smaller than the above value, 
a shell flash occurs (bottom panel).
}
\label{evol_m135}
\end{figure}

\subsection{Stability Line}
\label{section_stability}

To obtain the stability boundary of the accretion rate,
we start the evolution calculations from an equilibrium model 
whose accretion rate lies well within the stable range.
Then, we decrease the accretion rate slowly until a shell flash occurs. 
Figure\,\ref{evol_m135} depicts examples for $M_{\rm WD}=1.35~M_\sun$.
The initial model has a mass accretion rate of $\dot M=4.5 \times 10^{-7}
~M_\sun$~yr$^{-1}$.  In the top panel, the accretion rate $\dot M$
is reduced to $3.5 \times10^{-7}~M_\sun$~yr$^{-1}$. 
The hydrogen shell burning is still stable at this accretion rate,
and the nuclear luminosity $L_{\rm n}$ remains nearly constant.  

In the middle panel, $\dot M$ is decreased to 
$3.4\times10^{-7}~M_\sun$~yr$^{-1}$.
The nuclear luminosity $L_{\rm n}$ oscillates,  but the amplitude is bounded. 
If $\dot M$ is further decreased to $3.3\times10^{-7}~M_\sun$~yr$^{-1}$,
the nuclear luminosity $L_{\rm n}$ starts decreasing, 
and after a certain mass is accreted, a shell flash occurs (bottom panel). 
The stability of hydrogen shell burning changes at  
$\dot M=3.4\times10^{-7}~M_\sun$~yr$^{-1}$ for $M_{\rm WD}=1.35~M_\sun$.
We regard this accretion rate as the boundary of stability. 

We note that the appearance of oscillations in $L_{\rm n}$
for the models with accretion rates around the stability limit is consistent 
with the prediction made by \citet{pac83} from a linear stability analysis for
a simplified plane-parallel one-zone model.
He found that the eigenvalues of the linear analysis are always complex
near the transition from stability to instability because of the 
similarity between the thermal and nuclear timescales, 
which means that the temporal variations
in the perturbations are oscillatory.

\begin{figure}
\epsscale{1.2}
\plotone{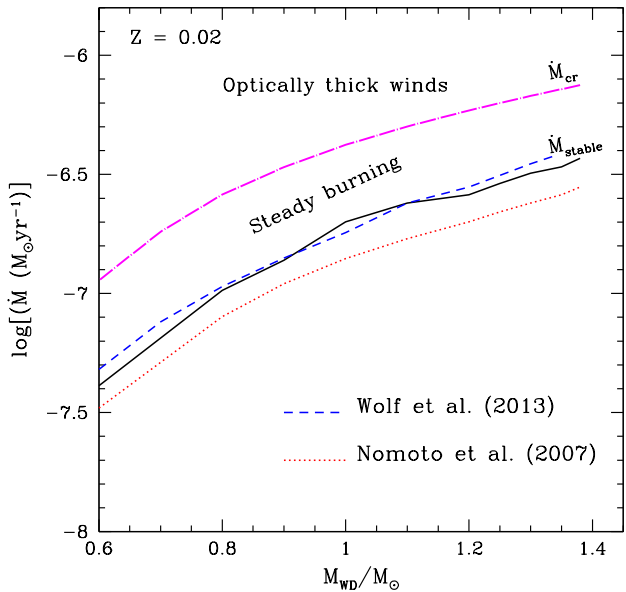}
\caption{
Comparison of stability line calculated in this work (black solid line) 
with those of \citet{nom07} (red dotted line) 
and \citet{wol13} (blue dashed line) in the $M_{\rm WD} - \dot M$
diagram.  Other symbols/lines are the same as in Figure 
\ref{revised_Nomoto_diagram}.
See Section \ref{section_stability} for details.
\label{mmdot_compare}}
\end{figure}

Figure \ref{mmdot_compare} compares the stability line we obtained 
(black solid line) with those of \citet{nom07} (red dotted line) and 
\citet{wol13} (blue dashed line).  \citet{nom07} obtained 
the stability line from a linear perturbation analysis of 
equilibrium models, 
whereas \citet{wol13} obtained it based on their evolution calculations.
The two lines based on time evolution calculations  
agree well with each other,
whereas that based on the linear perturbation analysis lies
below them. 

The reason for the difference is not clear.  Although the 
linear stability analysis of \citet{nom07} 
includes only thermal effects, and hence
only real eigenvalues are expected, \cite{pac83} showed that the monotonic
thermal stability transition occurs at the same accretion rate
as oscillatory instability. The difference might stem from small
differences in the structures between the equilibrium models and 
time evolution models.

\subsection{Shortest Recurrence Periods}
\label{min_period}

Now that we have obtained $M_{\rm ig}$ for various $M_{\rm WD}$
and $\dot M$, we can approximate the recurrence periods of novae.
After a nova outburst sets in, the envelope mass of the WD decreases as its
surface temperature increases.
Hydrogen burning stops at point E  in Figure \ref{flash} 
when the envelope mass reaches a local minimum, $M_{\rm env}^{\rm min}$.  
This hydrogen-rich matter remains unburnt until
the next flash occurs; during this time, an amount of mass equal to 
$M_{\rm ig}-M_{\rm env}^{\rm min}$ is accreted.
We estimate the recurrence period as 
$t_{\rm rec}=(M_{\rm ig}-M_{\rm env}^{\rm min})/\dot M$ because the time 
after ignition to point E is much shorter than the accretion time. 
We show in Section \ref{multi_cycles} that the 
recurrence periods determined in this way are
consistent with the evolution calculations of several flashes.

\begin{figure}
\epsscale{1.2}
\plotone{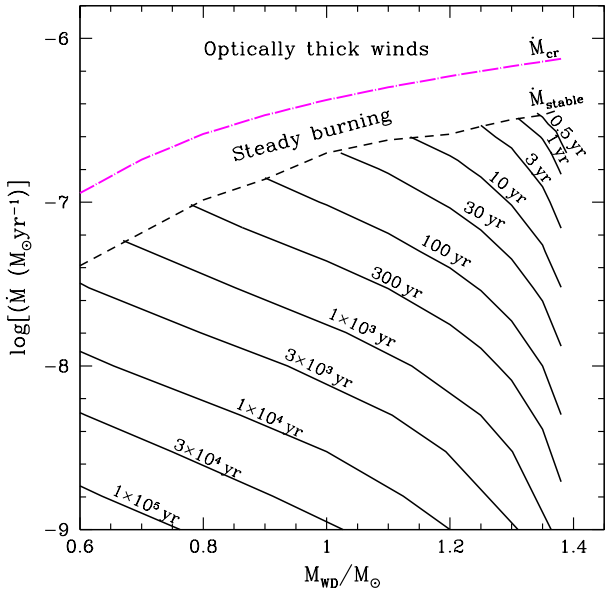}
\caption{
Recurrence period $t_{\rm rec}$ of novae on the $M_{\rm WD}-\dot M$ plane.
We plot the loci of the equi-recurrence periods of novae (black solid lines). 
Other symbols/lines are the same as in Figure \ref{revised_Nomoto_diagram}.
\label{mmdot_rtime}}
\end{figure}

Figure \ref{mmdot_rtime} shows the contours of the recurrence period
on the $M_{\rm WD}-\dot M$ plane.  The recurrence period of 1 yr 
corresponding to the nova M31N 2008-12a is obtained for WDs with
$(\dot M, ~M_{\rm WD})=(3.3 \times 10^{-7}~M_\sun$~yr$^{-1}, 
~1.31~M_\sun)$, $(2.4 \times 10^{-7}~M_\sun$~yr$^{-1}, ~1.35~M_\sun$),
and $(1.5 \times 10^{-7}~M_\sun$~yr$^{-1}, ~1.38~M_\sun)$. 
In multicycle nova outbursts calculated  by \citet{pri95}, 
a 1 yr recurrence period corresponds to massive WDs such 
as $(\dot M, M_{\rm WD}) = (\sim 10^{-7}~M_\sun$~yr$^{-1}, 1.4~M_\sun)$.
Wolf et al. (2013) and \citet{tan14} also obtained a 1 yr period for 
$(\dot M, M_{\rm WD}) = (3.4 \times 10^{-7}~M_\sun$~yr$^{-1}, 
1.30~M_\sun)$  and ($1.7 \times 10^{-7}~M_\sun$~yr$^{-1}, 1.36~M_\sun)$. 
Our values are consistent with these works.

For a given $\dot M$,
the recurrence period is shorter for a larger $M_{\rm WD}$.  
For a given $M_{\rm WD}$, the recurrence period is shorter 
for a higher $\dot M$.  The latter is limited by the stability line, 
as indicated by the dashed line in Figure \ref{mmdot_rtime}, which
gives a lower bound of the recurrence period for a given $M_{\rm WD}$. 
We should note that for the accretion rates as high as 
$\dot M > 10^{-7}~M_\sun$~yr$^{-1}$ encountered here, 
the accreting C+O WD becomes an SN Ia when $M_{\rm WD}$ reaches 
$1.38~M_\sun$ because carbon deflagration is ignited 
at the center of the WD \citep{nom82, nom84ty, hil00}.
If the accreting WD is an O+Ne+Mg WD, 
electron capture triggers a collapse when $M_{\rm WD} = 1.38~M_\sun$ 
\citep{nom84, can90}.
Therefore, the {\sl shortest} recurrence period of novae is obtained for 
the $1.38~M_\sun$ WD with an accretion rate of 
$3.6 \times 10^{-7}~M_\sun$~yr$^{-1}$, and is found to be 
about 2 months (see Figure \ref{mig_rect_1.38_X7X1} below).

\section{DISCUSSION}
\label{sec_discussion}

\subsection{Comparison with Other Works: Under the Stability Line}

\citet{pri95} calculated the shell flashes for various WD masses and 
accretion rates. For $\dot M =1 \times 10^{-7} ~M_\sun$~yr$^{-1}$ with 
a core temperature of $T_{\rm c}$ = $5 \times 10^7$ K, they obtained 
the WD mass, accreted mass, and recurrence period as
($1.0~M_\sun$, $8.30 \times 10^{-6}~M_\sun$, 83~yr), 
($1.25~M_\sun$, $1.96 \times 10^{-6}~M_\sun$, 19.6~yr), and
($1.4~M_\sun$, $8.09 \times 10^{-8}~M_\sun$, 0.809~yr).
For the same accretion rate of $1 \times 10^{-7}~M_\sun$ yr$^{-1}$,
\citet{ida13} reported values of 
($1.0~M_\sun$, $9 \times 10^{-6}~M_\sun$, 100~yr), 
($1.25~M_\sun$, $1.5 \times 10^{-6}~M_\sun$, 16~yr), 
($1.35~M_\sun$, $3.5 \times 10^{-7}~M_\sun$, 3.7~yr), and
($1.4~M_\sun$, $5 \times 10^{-8}~M_\sun$, 0.5~yr). 
Our models yield 
($1.0~M_\sun$, $1.3 \times 10^{-5}~M_\sun$, 95 yr), 
($1.25~M_\sun$, $2.3 \times 10^{-6}~M_\sun$, 17 yr ), 
($1.35~M_\sun$, $5.5 \times 10^{-7}~M_\sun$, 4.2 yr), and
($1.38~M_\sun$, $2.4 \times 10^{-7}~M_\sun$, 1.8 yr). 
These values are more or less consistent with each other. 
In our model, the WD is heated by gravitational 
energy release due to high mass accretion rates.  As a result, the WD radius
(i.e., the bottom of the envelope) should be slightly larger than 
those of \citet{ida13} and \citet{pri95}. Thus, our envelope 
mass is larger than theirs because of the smaller surface gravity of the WD.

Near the stability line, our ignition mass
$M_{\rm ig}$ is very consistent with the results of Wolf et al. (2013), 
as indicated by the agreement of the two stability lines 
in Figure \ref{mmdot_compare}. 
For high mass accretion rates, Wolf et al.'s models develop a hot helium 
layer below the accreted hydrogen-rich envelope after the WD experiences
several shell flashes. Wolf et al. claimed that the presence of 
a hot helium layer is essential for nova outbursts 
of short recurrence periods because it acts  
as a heat reservoir even if the WD is cool.  
Our equilibrium model has no helium layer but has a hot WD core, which
is heated by compressional heating due to high mass accretion rates.
For example, the maximum temperature of our $1.0~M_\sun$ WD model 
with $\dot M=1.0\times10^{-7}~M_\sun$~yr$^{-1}$ reaches $\log T$ (K)=8.07
at the outer part of the WD core,
which is comparable with that of Wolf et al.'s helium layer. 
Therefore, our models and Wolf et al.'s have a similar inner boundary 
condition in view of the heat reservoir at the bottom of the hydrogen-rich
envelope, and thus give very consistent results.

\begin{figure}
\epsscale{1.2}
\plotone{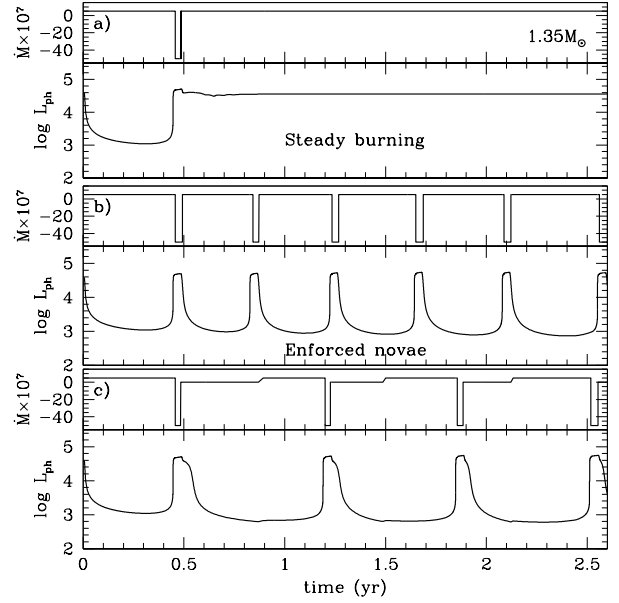}
\caption{
Evolutions of photospheric luminosity $\log L_{\rm ph}$ 
(in  units of $L_\sun$) of a $1.35~M_\sun$ WD
for an accretion rate of $\dot M = 5\times 10^{-7}~M_\sun$~yr$^{-1}$,
above the stability line.  Each calculation started with about half
of the equilibrium envelope mass; 
therefore, the envelope underwent a first shell flash at about 0.45 yr.
To avoid numerical difficulty, we assumed an artificial wind mass loss of
$\dot M = -5\times 10^{-6}~M_\sun$~yr$^{-1}$ when the photosphere 
expands to $\log R_{\rm ph}/R_\sun > -1.7$.
In the upper part of each panel, we plot our control 
mass accretion/mass loss rate in units of $10^{-7}~M_\sun$~yr$^{-1}$.
(a) Upper panel: after the first shell flash, 
we stopped the wind mass loss and restarted
mass accretion when the photospheric radius shrank to 
$\log R_{\rm ph}/R_\sun =-2.0$, where the envelope had not yet 
reached point E in Figure \ref{flash}.  Hydrogen burning
was occurring (steady state).  The envelope remained on the stable branch
in Figure \ref{flash}.
(b) Middle panel: after the first shell flash,
we stopped the wind mass loss and restarted 
mass accretion when the photospheric radius shrank to 
$\log R_{\rm ph}/R_\sun =-2.4$, where the envelope had already reached
point E in Figure \ref{flash} and began to fall toward point A.
The envelope repeated a limit cycle (enforced novae).
(c) Bottom panel: after the first shell flash,
we stopped the wind mass loss when the photospheric radius shrank to 
$\log R_{\rm ph}/R_\sun =-2.0$, but did not restart mass accretion, 
i.e., $\dot M=0$, until the photospheric luminosity decreases to 
$\log L_{\rm ph}/L_\sun < 2.8$.  The envelope gradually moved toward  
and reached point E in Figure \ref{flash}.  After the envelope fell 
toward point A, we restarted mass accretion.  The envelope again underwent
a shell flash (enforced nova).  The recurrence period is longer than
in the middle panel.
}
\label{evol_m135_novae_3cases}
\end{figure}

\subsection{Shell Flashes above the Stability Line: Artifact Flashes}
\label{enforced_novae}

Although the stability line of shell flashes 
in Figure \ref{revised_Nomoto_diagram} has long been established
\citep{ibe82, nom82, nom07, pac78, sie75, sie80, sio79, wol13}, 
some numerical calculations showed 
repeated shell flashes for accretion rates even above the stability line 
\citep{kov94,pri95,sta12,ida13}. In these cases, the recurrence period
is much shorter than our minimum value,  
and the light curves show a very high duty cycle.  
For example, \citet{kov94} reported the calculation
of several successive shell flashes on a $1.4~M_\sun$ 
WD with $\dot M=1 \times 10^{-6}~M_\sun$~yr$^{-1}$. 
The bolometric luminosity varies between 
$\log (L_{\rm bol}/L_\sun) \sim 3.7$ and $\sim 4.8$, 
and the bright stage lasts 10 days,
almost half of the recurrence period of 20 days.  
\citet{ida13} obtained successive nova outbursts for a $1.0~M_\sun$ WD with 
$\dot M=1\times10^{-6}~M_\sun$~yr$^{-1}$. 
The bright stage lasts 6 yr, 60\% of the total recurrence period of 10 yr. 
Both of these calculations show a long lasting high-luminosity phase during
the relatively short recurrence period.  Such light curves do not resemble
those of recurrent novae, which show much shorter duty cycles. 
 
The accretion rate of $\dot M=1 \times 10^{-6}~M_\sun$~yr$^{-1}$ 
corresponds to the region above the stability line
in Figure \ref{revised_Nomoto_diagram}, 
in which hydrogen shell burning should be stable, as discussed
in Section \ref{section_limitcycle} (Figure \ref{flash}).
Nevertheless, these papers reported successive nova outbursts.
This is because they stopped accretion during the shell flashes. 
Without a supply of hydrogen (nuclear fuel), a nova evolves toward point E
in Figure \ref{flash}, where it immediately falls to point A. 
These authors then restarted accretion.  After some time, 
the next shell flash begins, and it goes up from point B to point C. 
In this way, these authors switched mass accretion on and off 
to run the limit cycle. In other words, these outbursts are 
``enforced novae,'' and it is not known whether such an on/off switch works
in nova outbursts or in other accreting binaries. 

Figure \ref{evol_m135_novae_3cases} shows examples of our numerical
simulations for such enforced novae for a $1.35~M_\sun$ WD with
$\dot M = 5\times 10^{-7}~M_\sun$~yr$^{-1}$,
which is above the stability line.  
We show three cases for different on/off switches for mass accretion.
Each calculation started with about half of the equilibrium envelope mass;
therefore, the envelope underwent the first shell flash at about 0.45 yr.
To avoid numerical difficulty, we assumed an artificial wind mass loss of
$\dot M = -5\times 10^{-6}~M_\sun$~yr$^{-1}$ when the photosphere 
expands greatly ($\log R_{\rm ph}/R_\sun > -1.7$).  
(a) Upper panel: after the first shell flash, 
we stopped the wind mass loss when the photospheric radius shrank to 
$\log R_{\rm ph}/R_\sun =-2.0$, 
where the envelope did not yet reach point E in Figure \ref{flash},  
and immediately restarted mass accretion.  
 Hydrogen burning was occurring (steady state).  
The envelope remained on the stable branch
in Figure \ref{flash}.  In the upper part of each panel, we plot our control 
mass accretion/mass loss rate in units of $10^{-7}~M_\sun$~yr$^{-1}$.
(b) Middle panel: after the first shell flash, we stopped the wind mass loss
and immediately restarted mass accretion when the photospheric radius
shrank to $\log R_{\rm ph}/R_\sun =-2.4$, which is a bit later 
than in (a).  The envelope had already passed through point E 
in Figure \ref{flash} and fell toward point A. 
After we restarted mass accretion,  the envelope repeated a limit cycle. 
This is the enforced nova outburst.
(c) Bottom panel: after the first shell flash,
we stopped the wind mass loss when the photospheric radius shrank to 
$\log R_{\rm ph}/R_\sun =-2.0$, the same time as in (a),
but did not restart mass accretion, i.e., $\dot M=0$, 
until the photospheric luminosity further decreased to 
$\log L_{\rm ph}/L_\sun < 2.8$.  We restarted mass accretion 
after the envelope fell from point E toward point A 
in Figure \ref{flash}.  The envelope again underwent
a shell flash (enforced nova).  The recurrence period was longer than
in the middle panel because we delayed the restart of mass accretion.
In this way, we can control the recurrence period by changing the 
time at which mass accretion is restarted.  The above three cases clearly show 
that shell flashes above the stability line are just artifacts that stem
from the numerical treatment of the mass accretion on/off switch. 

There is some indication that accretion does not stop in real binaries. 
When the accretion rate is between the two lines of $\dot M_{\rm cr}$ and 
$\dot M_{\rm stable}$ in Figure \ref{revised_Nomoto_diagram}, 
the WD burns hydrogen at the same rate as the accretion rate
and stays somewhere between points D and E in Figure \ref{flash}.  
These binaries correspond to steady SSSs such as CAL 83,
1E0035.4$-$7230, and RX J0925.7$-$4758 \citep[e.g.,][]{vdh92, kah97, nom07}.
In these binaries, the WD is considered to accrete matter from the disk 
and to emit X-rays from the other WD surface area.  

If the accretion rate is larger than $\dot M_{\rm cr}$, the star stays 
somewhere to the right of point D in Figure \ref{flash},
where optically thick winds occur.  The mass accretion rate is
balanced with the sum of the nuclear burning rate and the 
wind mass loss rate.  In this case, the WD accretes matter 
from the disk and ejects excess matter in the wind from the other regions.  
RX~J0513.9$-$6951 and V~Sge are considered to be such objects 
\citep{hac03a, hac03b}. 

\begin{figure}
\epsscale{1.2}
\plotone{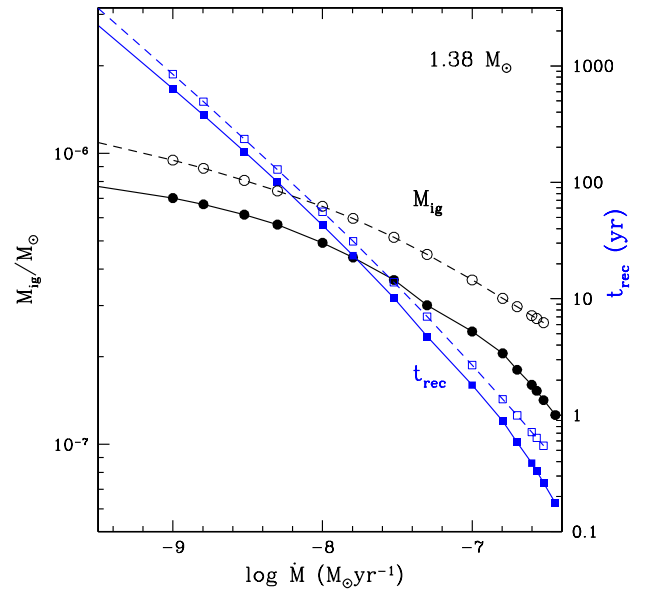}
\caption{
Recurrence period and ignition mass of novae on a $1.38~M_\sun$ WD.
Black lines/symbols represent ignition masses (left vertical axis).  
Blue lines/symbols represent recurrence period
$t_{\rm rec}$ (right vertical axis). 
Solid lines with filled symbols indicate 
the case of $X=0.7$, $Y=0.28$, and $Z=0.02$, 
whereas dashed lines with open symbols denote the case
of $X=0.1$, $Y=0.88$, and $Z=0.02$.
The minimum recurrence period is about two months for 
$X=0.7$, $Y=0.28$, and $Z=0.02$.
\label{mig_rect_1.38_X7X1}}
\end{figure}

\subsection{Effects of Chemical Composition} 
\label{sec_composition}
The ignition mass depends on the chemical composition of the envelope
as well as the WD mass and mass accretion rate.
Figure \ref{mig_rect_1.38_X7X1} compares the ignition mass and 
recurrence period for $X=0.1$, $Y=0.88$, and $Z=0.02$ with those for 
the solar composition.  The ignition mass for $X=0.1$ is systematically
larger than that for $X=0.7$. The nuclear burning rate due to the $pp$-chain 
is proportional to $X^2$, and the rate of the CNO cycle is proportional
to $X$.  For smaller $X$, unstable nuclear burning ignites at a higher
temperature $T_{\rm c}$, which results in a more massive envelope. 
This will be useful in binary evolution calculations toward SNe~Ia, 
in which a WD accretes He-rich matter, 
as suggested for U Sco type binaries \citep{hku99}.

\begin{figure}
\epsscale{1.2}
\plotone{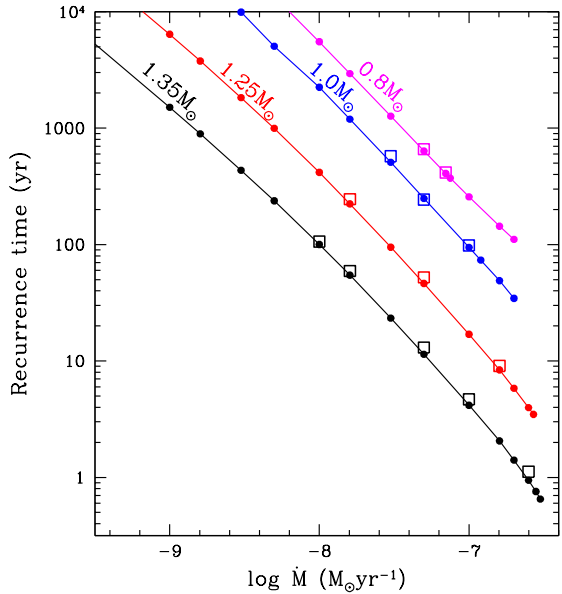}
\caption{
Recurrence period vs. mass accretion rate
for 0.8, 1.0, 1.25, and $1.35~M_\sun$ WDs.
Solid lines with small filled circles correspond to the recurrence
time calculated from the envelope mass at the ignition of the first flash,
as in Figure \ref{mmdot_rtime},
and large open squares are the recurrence periods obtained by calculating
many cycles of flashes until the repetition became nearly a limit cycle.
See Section \ref{multi_cycles} for more details.
\label{mdot_rect_evol_rv}}
\end{figure}

\subsection{Recurrence Periods after Multicycle Nova Outbursts}
\label{multi_cycles}

The ignition mass shown in Figure \ref{revised_Nomoto_diagram}
is obtained from the first shell flash of the model evolved 
from the equilibrium model.  The recurrence time of nova outbursts 
in Figure \ref{mmdot_rtime} is estimated from the ignition mass 
and the minimum envelope mass for each WD mass.
To verify these estimates, we calculated several consecutive
shell flashes starting from the equilibrium model for selected cases.

Figure \ref{TcMenvM138X70.s} shows five successive cycles
of shell flashes for $M_{\rm WD}=1.38~M_\sun$ and
$\dot M=2\times10^{-7}M_\sun$~yr$^{-1}$.
These five tracks are very similar.  In general,
if we start from an arbitrary initial thermal condition of the WD,
for example, a cold WD, the strength and recurrence period of the shell flashes
would vary from flash to flash and gradually approach a final limit cycle
\citep[see e.g.][]{ibe82, kov94}.  To avoid a lengthy calculation,
as the initial model we adopted the equilibrium model, which is 
a good approximation of the long time-averaged evolution of WDs, 
as shown in Figure \ref{TcMenvM138X70.s} and also the discussion below.   

Figure \ref{mdot_rect_evol_rv} shows the recurrence periods
based on the ignition mass of the first flash (filled circles)
and obtained from the last shell flash (open squares) after 
several consecutive nova outbursts. The shell flashes were calculated
by assuming mass loss in the expanded stages (Section \ref{sec_evolution}).
The calculations were continued until the repetition of the flashes
became nearly a limit cycle.
The calculations for longer recurrence times are very difficult because
the shell flashes are very strong.
For this reason, the recurrence periods obtained from multicycle
calculations are limited to about a thousand years.

In the range where two types of recurrence periods are available,
the periods agree with each other, as seen in Figure\,\ref{mdot_rect_evol_rv}.
Actually, these two recurrence periods agree within 10\%, suggesting that
our estimates of the recurrence periods based on the first-cycle
ignition mass and the minimum envelope mass given in Section \ref{min_period}
(Figure\,\ref{mmdot_rtime}) are reasonable.

\section{CONCLUSIONS}
\label{conclusions}

The main results are summarized as follows.

\noindent
1. We proposed a physical mechanism that leads to a finite minimum 
recurrence period of novae.

\noindent
2. We calculated the ignition masses for various WD masses and
mass accretion rates.  We determined that the shortest recurrence period
of novae is about two months for a non-rotating $1.38~M_\sun$ WD 
with a mass accretion rate of $3.6 \times 10^{-7}~M_\sun$~yr$^{-1}$. 

\noindent
3. A 1 yr recurrence period of a nova is possible 
only for very massive WDs ($M_{\rm WD} \gtrsim 1.3~ M_\sun$) and 
very high mass accretion rates, 
e.g., $\dot M =3.3 \times 10^{-7}~M_\sun$~yr$^{-1}$ for a $1.31~M_\sun$ WD, 
$2.4 \times 10^{-7}~M_\sun$~yr$^{-1}$ for a $1.35~M_\sun$ WD, 
and $1.5 \times 10^{-7}~M_\sun$~yr$^{-1}$ for a $1.38~M_\sun$ WD.  

\noindent
4. We revised our stability limit of hydrogen shell burning
\citep{nom07}, which is useful for binary evolution calculations toward 
SNe Ia.

\acknowledgments

We thank the anonymous referee for useful comments 
that improved the manuscript.  
This research was supported in part by 
Grants-in-Aid for Scientific Research (22540254, 23224004, 23540262, 24540227, 
and 26400222) from the Japan Society for the Promotion of Science 
and by the WPI Initiative, MEXT, Japan.

\end{document}